\documentclass{aa}
\usepackage{graphicx}
\usepackage{natbib}
\bibpunct{(}{)}{;}{a}{}{,}

\begin{document}

\titlerunning{Microlensing towards LMC}
\title{Microlensing towards LMC:\\a study of the LMC halo contribution}

\authorrunning{Calchi Novati et al.}
\author{S.~Calchi Novati \inst{1,2,4} \and F.~De Luca \inst{1,2} 
\and Ph.~Jetzer \inst{1} \and G.~Scarpetta \inst{2,3,4}}

\institute{
Institute for Theoretical Physics,  University of Z\"urich, 
Winterthurerstrasse 190, CH-8057 Z\"urich, Switzerland \and
Dipartimento di Fisica ``E. R. Caianiello'', 
Universit\`a di Salerno,
Via S. Allende, I-84081 Baronissi (SA), Italy \and
International Institute for Advanced Scientific Studies, 
Vietri sul Mare (SA), Italy \and
Istituto Nazionale di Fisica Nucleare, sez. Napoli, Italy
}
\date{Received/ Accepted}

\abstract{
We carry on a new analysis of the sample of MACHO microlensing candidates
towards the LMC.
Our main purpose is to determine the lens population
to which the events may belong. We give particular
emphasis to the possibility of characterizing
the Milky Way dark matter halo population
with respect to the LMC one. Indeed,
we show that only a fraction of the events
have characteristics that
match those expected  for lenses belonging 
to the MACHO population of the Milky Way halo.
This suggests that this component cannot explain all the candidates.
Accordingly, we challenge the view that the dark matter 
halo fraction of both the Milky Way
and the LMC halos are equal, and indeed we show that,
for a MACHO mass in the range 0.1-0.3 M$_\odot$, the LMC halo
fraction can be significantly larger than the Milky Way one.
In this perspective, our main conclusion is
that up to about half of the observed events
could be attributed to the LMC MACHO dark matter halo. 
}
\maketitle

\section{Introduction} 
Since the original proposal of \cite{pacz86},  microlensing has shown 
to be a powerful tool
for the investigation of the MACHO composition of the galactic halos. 
The microlensing surveys
towards the LMC and the SMC probed the existence
of compact halo objects along these lines of sight,
however, the assessment of these results with respect to the actual 
fraction of the Galactic halo in form of MACHOs is still highly debated.

The MACHO collaboration reported the detection
of 13-17 microlensing candidates towards the LMC \citep{MACHO00},
arguing in favour of a MACHO halo fraction
of $\sim 20\%$ of objects of $\sim 0.5\,\textrm{M}_\odot$, 
and estimating a microlensing optical depth towards the LMC of 
$\tau = 1.2^{+0.4}_{-0.3}\times 10^{-7}$. The reported microlensing 
rate towards the LMC significantly exceeds the expected one 
from known visible components of our Galaxy.
Further analysis mainly confirmed these conclusions
\citep{bennett05b}. On the other hand, the EROS collaboration,
out of observations towards both the LMC and the SMC,
put more and more lower \emph{upper} limits
on the MACHO contribution to galactic halos \citep{eros00,eros03,eros05},
that are no longer compatible with the MACHO results. 

These questions have been addressed
also by microlensing surveys towards M31 \citep{agape97},
and first evidences for a MACHO contribution along
this line of sight have been reported \citep{novati05},
although challenged in \cite{mega06}.
Overall, the picture remains unclear, in particular the problem
of the nature and the location of the observed events. 
For example, \cite{green02} have cast doubts on 
the interpretation of the microlensing data towards LMC as due to 
a dominant lens population made by MW halo MACHOs. 
Indeed they show,  at high level of confidence, that the distribution 
of the duration of the 
observed microlensing events is significantly narrower 
than what is expected from a standard halo lens population.

We have addressed some of these questions already in \cite{jetzer02} and 
\cite{mancini04} (hereafter respectively Paper I and Paper II).
A main issue in both works is that the microlensing events towards the LMC,  
observed  by the MACHO collaboration,  do not belong necessarily all 
to the same lens population.
In particular in Paper II we have considered the issue of self lensing
in the framework of the \cite{vdmarel02} picture of the
luminous components of the LMC. The main conclusion,
based both on the predicted number and characteristics
of self-lensing events, was that self lensing alone cannot
explain all the observed events.

In the present work we extend the analysis by fully considering
the LMC and MW dark matter halo MACHO lens populations. 
The main question we want to address 
is whether and to which extent  events due to the lens population 
residing in the LMC halo can contribute to the observed rates.
This issue was first proposed and discussed in \cite{gould93}. 
Here we consider
again the question taking into account the results 
of the MACHO collaboration 
and the most recent results on the modeling of the LMC.

The paper is organised as follows. In Sect. \ref{sec:models}
we review the models of the LMC and MW components we consider.
In Sect. \ref{sec:rate} we discuss the evaluation 
of the microlensing rate and present the results
for the expected number and duration of the microlensing events.
In Sect. \ref{sec:lmc-res} we carry out
our new analysis and present
our main conclusions on the LMC MACHO halo contribution,
and in Sect. \ref{sec:the-end} we present our conclusions.

\section{Models} \label{sec:models}

We consider the LMC as composed by a luminous part,
the bar and the disc, plus a stellar and a dark matter halo. We also include
the MW dark matter halo, but we do not include
the MW disc and spheroid populations. 
These components have already been shown to give
smaller contributions than the LMC self lensing \citep{MACHO00},
which we instead include. Accordingly, we exclude
from our analysis the only one event whose lens,
upon a direct search, has been acknowledged
to be part of these components.

For the structure and dynamics of the luminous components 
of the LMC we follow closely Paper II.
Out of the analysis of \cite{vdmarel02}, who derive
their results on the assumption that the carbon star population 
is representative of the bulk of the LMC disc stars, 
we take up the results on the 
LMC disc intrinsic ellipticity, vertical thickness, line-of-sight velocity
dispersion and rotation curve and the coincidence
of the centre of mass of the disc and the bar components.
We assume the following values for the bar and disc components 
$M_{\mathrm{bar}}+M_{\mathrm{disc}}=(2.7\pm 0.6) \times 10^{9}\,
\mathrm{M}_{\odot}$ \citep{vdmarel02}, and
$M_{\mathrm{bar}}=1/5\,M_{\mathrm{disc}}$ \citep{gyuk00},
that we consider to be both centered at 
$\alpha,\delta$ = $5^\mathrm{h} 27.6^\mathrm{m} \pm 3.9^\mathrm{m}, 
- 69.87^{\circ} \pm 0.41^{\circ}$ (J2000) 
at a distance from us of $D_{0} = 50.1 \pm 2.5 \, \mathrm{kpc}$ \citep{vdmarel02}.
We use the same density star distribution than in Paper II,
characterized by a vertical distribution for the exponential disc
described by a sech$^2$ function and a boxy bar, with a gaussian 
profile along the major axis and the section described 
by a $\exp({-r^4})$ function.
The vertical structure of the LMC has been recently the object
of great debate (for a discussion see \citealt{vdmarel04}).
We have considered this issue, with respect to the expected
self lensing signal, in Paper II.  
In the present analysis, where we  focus on the contribution of the two halos,
we do not enter such discussion and consider only
the configuration with coplanar disc and bar.

The presence of a significant LMC stellar halo population 
is a matter of debate \citep{minniti03,alves04,gallart04}.
In the present analysis we include the contribution
of such a possible component following \cite{alves04}
who proposes a spherically symmetric spheroid with density profile
\begin{equation} \label{eq:rhoh}
\rho=\rho_{\mathrm{0}}\,\left(1+\frac{{R}^{2}}{a_{\mathrm{C}}^{2}}\right)^{-k},
\end{equation}
with $k=3/2$, central density $\rho_{0} = 6.3 \times 10^{6}\,
\mathrm{M}_{\odot}\,\mathrm{kpc}^{-3}$
and core radius $a_{\mathrm{C}}=1.42\,\mathrm{kpc}$ for a total mass, 
within 8.9 kpc,
of $0.35\times 10^9 \mathrm{M}_{\odot}$ somewhat smaller than that of the bar. 
The optical depth profile of this component  shows a near-far
asymmetry due to the LMC disc inclination whose overall shape recalls
that of the optical depth profile of the LMC MACHO halo component
(Paper II, Figure 4), with a maximum value around $0.9\times 10^{-8}$
reached in correspondence of the field MACHO 6.

Following \cite{vdmarel02},
who present observational evidences based
on the rotation curves, we include a significant 
LMC dark matter halo component.
We assume a total LMC mass of $8.7\times10^{9}\mathrm{M}_{\odot}$
within 8.9 kpc with a truncation radius of 15 kpc \citep{vdmarel02},
that we consider to include also the contribution from the stellar halo.
We assume a spherical isothermal model ($k=1$ in Eq. \ref{eq:rhoh}) with 
a core radius of $a_{\mathrm{C}} = 2\,\mathrm{kpc}$ \citep{MACHO00}.

This spherical symmetric configuration might not be a realistic 
description of this component
given that the dynamical environment of the LMC can induce 
tidal distortions and disruptions
especially in the outer parts. To take into account this issue, in Paper II 
we have compared the LMC halo optical depth profiles for a spherical 
and an elliptical configuration (Fig. 4 and 5 respectively):
the overall shape is similar even if in the latter case the maximum 
value rises by about 20\% 
and the near-far asymmetry is enhanced. However, as we lack any strong constraint,
we prefer not to introduce  a further parameter in the present analysis, therefore 
we consider only the spherical configuration, in view also of the possibility 
to carry out a more direct comparison with previous works.

In this same perspective for the MW dark matter halo we consider 
the ``standard'' isothermal profile with a core radius of 5 kpc, local density
$7.9\times 10^{6}\, \mathrm{M}_{\odot}\,\mathrm{kpc}^{-3}$
and a distance from the Galactic centre of 8.5 kpc \citep{MACHO00}.

\section{The microlensing rate} \label{sec:rate}

The main tool of investigation we use is  
$\frac{\mathrm{d}\Gamma}{\mathrm{d}\,T_{\mathrm{E}}}$, the
differential rate of microlensing events with respect to the
Einstein time $T_{\mathrm{E}}$ \citep{derujula91,griest91,roulet97}.
This allows  us to make predictions on the timescale,  
the  number and the spatial distribution of the  expected events, 
that we can compare with the corresponding observed quantities.  
With respect to the self--lensing configuration we have analysed in Paper II 
(section 4.2), we can no longer adopt the useful approximation 
${D_{ol}\over D{os}}\equiv x \approx 1$.  
Moreover we now have to take fully into account the bulk velocity of 
the LMC components and the relative motion between the LMC and the 
MW \citep{vdmarel02}.  

The source stars belong to the luminous components, disc or bar, of the LMC, 
whereas the lenses can belong either to the LMC or the MW halo.  
We assume  an isotropic Maxwellian profile\footnote{The Maxwellian profile of the
velocity distribution is the first term of a series expansion in
terms of Gauss--Hermite moments \citep{vdmarel93,gerhard93}. See
Section 3.2 of Paper I.} for the velocity distribution for both lenses and sources.
For the flattened LMC luminous components
this is a rough approximation, still,
it  gives a fair description of the average properties of these populations,
that we consider to be sufficient in the present framework.

For the LMC disc component we consider the rotational
velocity as in \cite{vdmarel02} with $\sigma=20.2\,\mathrm{km/s}$.
For the LMC bar stars we use a larger value
of the velocity dispersion than for the disc, 
$\sigma=24.7\,\mathrm{km/s}$ \citep{cole05}.
For both the LMC halo components, stellar and MACHO, we use
$\sigma=46\,\mathrm{km/s}$ \citep{vdmarel02,alves04} 
(we have tested that our results remain qualitatively unaltered
by changing this value up to 20\%);
for the MW halo,  $\sigma=155\,\mathrm{km/s}$.

The expression for the random motion velocity for 
the lenses reads\footnote{The velocity components
parallel to the line of sight are integrated out, 
and the subscript $\perp$ 
indicates the vectorial component in the plane orthogonal to the line of sight.}
\begin{equation}
\label{eq:randomVelocity}
{{\vec v}}_{{l}\perp} = {\hat{\vec v}}_{{l}\perp}+ x\, 
{{\vec{v}}}_{\mathrm{s}\perp}\ + {\vec A}_\perp\,,
\end{equation}
where ${\hat{\vec v}}_{{l}\perp}$ is the velocity
relative to the microlensing tube at position $x$, whose modulus is the ratio
between the Einstein radius and the Einstein time 
(${\hat{v}}_{{l}\perp}= R_{E}/T_{E}$),
${{\vec{v}}}_{\mathrm{s}\perp}$ the random component 
of the velocity of the sources.
All the bulk motions are included in ${\vec A}_\perp$, defined as
\begin{eqnarray}
{\vec A}_\perp&=&{{\tilde{\vec{v}}}_{\odot\perp}}+
\,x\,\left({\tilde{\vec{v}}}_{\mathrm{LMC}\perp}-
\,{{\tilde{\vec{v}}}_{\odot\perp}}+
\,{{\vec{v}}}_{\mathrm{s,drift}\perp}\right) -\nonumber \\  
&&\eta\, \left({\tilde{\vec{v}}}_{\mathrm{LMC}\perp} + 
{{\vec{v}}}_{l,\mathrm{drift}\perp}\right)
\end{eqnarray}
where $\eta = 0, \, 1$ for lenses in the Galaxy and in the LMC respectively,  
${{\vec{v}}}_{l,\mathrm{drift}\perp}$ is the drift velocity of the lens star 
belonging to the LMC disc (for the bar sources as well as
for the halo lenses we only consider a random motion component)
and a tilde over a vector indicates a quantity measured  
by an observer comoving with the MW centre.
In the self--lensing configuration, 
${\vec A}_\perp\approx 0$, and  Eq.   (\ref{eq:randomVelocity}) 
reduces to Eq. (13) of Paper II.

\begin{figure}[hbt!]
\resizebox{\hsize}{!}{\includegraphics[scale=0.8]{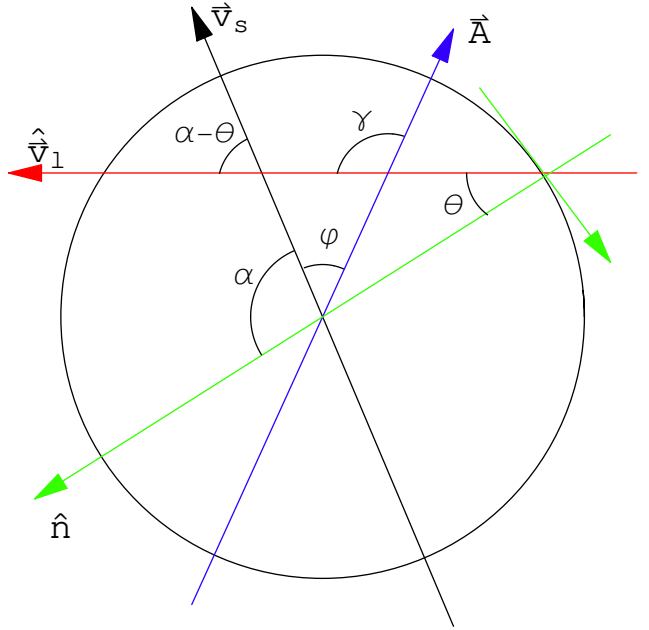}}
\caption{Scheme of a section of the microlensing tube with 
indicated the positions of the vectors and the angles involved. 
\label{fig:angoli}}
\end{figure}

We call  $\alpha$ the angle between the inner normal to the tube, $\hat {\vec n}$, 
and the source velocity ${{\vec{v}}}_{\mathrm{s}\perp}$;  
$\theta$ the angle between $\hat {\vec n}$ and ${\hat{\vec v}}_{{l}\perp}$, $\theta \in (-\pi/2,\pi/2)$;
$\varphi$ the angle between  ${{\vec{v}}}_{\mathrm{s}\perp}$ and ${\vec A}_\perp$; 
so that the angle between  ${{\vec v}}_{{l}\perp}$ 
and ${{\vec{v}}}_{\mathrm{s}\perp}$ 
is $\alpha - \theta$, 
and that between ${{\vec v}}_{{l}\perp}$ and ${\vec A}_\perp$ is 
$\gamma=\alpha +\varphi-\theta$. Both $\alpha$ and $\varphi$ vary 
in the range ($0,\, 2\pi$) (Fig. \ref{fig:angoli}). 

In the case of ${\vec A}_\perp=0$ all the integrations over the angular 
variables can be carried out analytically. This is now only possible for 
the $\theta$ variable. Altogether, after an analytical integration 
on the modulus of the source velocity, we are left with the following expression 
of the differential rate with respect to the Einstein time $T_{\mathrm{E}}$:
\begin{eqnarray}\label{eq:rate1}
\frac{\mathrm{d}\Gamma}{\mathrm{d}T_{E}} & = & \int_0^{2\pi}{\mathrm{d}\alpha} 
\int_0^{2\pi}{\mathrm{d}\varphi} \int_{d_{\mathrm{min}}}^{d_{\mathrm{max}}}
\mathrm{d}D_{os}\int_{{x_{\mathrm{min}}}}^{1} \mathrm{d}x
\int_{\mu_{\mathrm{min}}}^{\mu_{\mathrm{max}}} \mathrm{d}\mu \,\, 
\nonumber \\ 
&&
\frac{\mathrm{d}n_{l}}{\mathrm{d}\mu}\,\frac{D_{os} \, 
\rho_{l} \, \rho_\mathrm{s}\,{\hat v}_{{l}\perp}^{4}}{2\,\pi^{2} 
(\sigma^2_{l} + x^{2} \sigma^{2}_\mathrm{s}) \,{\cal N}} \,
\times\nonumber \\ 
&&
\Bigg\{ \exp\Big[{-\frac{A^{2}+{\hat v}_{{l}\perp}^{2}+2 A\,  
{\hat v}_{{l}\perp} \cos(\alpha+\varphi)} {2\,\sigma^{2}_{l}}}\Big]-
\nonumber \\ 
&&
\frac{\sqrt{\pi} \, x \, \sigma_\mathrm{s}\, \big({\hat v}_{{l}\perp} 
\cos(\alpha) +A \cos(\varphi)\big)}
{\sigma_{l} \sqrt{2\,(\sigma^{2}_{l}+x^{2} \sigma^{2}_\mathrm{s})}} 
\times\nonumber \\ 
&&
\exp\Big[-\frac{A^{2}+{\hat v}^{2}_{{l}\perp}+2 A\,  
{\hat v}_{{l}\perp} \cos(\alpha+\varphi)}{2\,(\sigma^{2}_{l}+x^{2} 
\sigma^{2}_\mathrm{s})}-
\nonumber \\ 
&&
\frac{x^2 \, \sigma_\mathrm{s}^2\,({\hat v}_{{l}\perp} 
\sin(\alpha)-A \sin(\varphi))^2}{\sigma_{l}^2 \,2\,(\sigma^{2}_{l}+
x^{2} \sigma^{2}_\mathrm{s})}
\Big] \times \nonumber \\ 
&&\Big(1-\mathrm{Erf} \big(\frac{x \sigma_\mathrm{s}\,({\hat v}_{{l}\perp} 
\cos(\alpha)+A\cos(\varphi))}{\sigma_{l} 
\sqrt{2(\sigma^{2}_{l}+x^{2} \sigma^{2}_\mathrm{s})}}\,
\big) \Big)\Bigg\} \;, 
\end{eqnarray}
where  the normalization factor ${\cal N}$ is the integral over the line of
sight of the sources
$$
{\cal N} = \int_{d_{\mathrm{min}}}^{d_{\mathrm{max}}}
\mathrm{d}D_{os}\,\rho_{\mathrm{s}}(D_{\mathrm{os}})\,,
$$
having assumed that the number of detectable stars varies 
with the distance as $D_{\mathrm{os}}^{-2}$. We define
$x_{\mathrm{min}} ={d_{\mathrm{min}}}/{D_{\mathrm{os}}}$,
$A=|\vec{A}|$; the integration limits along the line of sight,
$d_{\mathrm{min}}$ and $d_{\mathrm{max}}$, represent the distances
from the observer to the intersection  with the LMC tidal surface, 
for lenses in the LMC, whereas
$d_{\mathrm{min}}=0$ for lenses in the Galactic halo.
We use solar mass units, defining $\mu =
{\frac{M}{\mathrm{M}_{\odot}}}$, where $M$ is the lens mass.
For lenses belonging either to the disc or the bar
of the LMC, as well as for the stellar LMC halo contribution, 
we use the exponential function as in \cite{chabrier01} 
for the mass function  $\frac{\mathrm{d} n_{l}}{\mathrm{d} \mu}$, 
with integration limits $\mu_{\mathrm{min}}=0.08$ and $\mu_{\mathrm{max}}=1.5$.
If the lenses belong to the MW  halo or to the LMC halo, the would be MACHOs, we 
use a set of delta function with values going from $10^{-5}$ up to 10 M$_\odot$.

Finally, to compare with the results of the observations
we have to take into account the expression
for the detection efficiency, so that we obtain
\begin{equation} \label{eq:rate2}
\left({\frac{\mathrm{d}\Gamma}{\mathrm{d}T_{\mathrm{E}}}}\right)_{\varepsilon}
={\frac{\mathrm{d}\Gamma}{\mathrm{d}T_{\mathrm{E}}}}\,
\cdot\,{\cal E}(T_{\mathrm{E}})\,.
\end{equation}

To discuss the results of the MACHO collaboration
we use the efficiency as a function of $T_\mathrm{E}$ as in \cite{MACHO00}
for which an analytical approximation is given in Paper II.
We take into account the correction reported in \cite{bennett05b}.

\subsection{Expected number and duration} 

Following the prescription outlined in the previous section,
we evaluate the differential microlensing rate,
for each different lens population we consider,
along the lines of sight towards the MACHO fields.
In particular for each lens population we
calculate the expected number of events per field as
\begin{equation} \label{eq:nev}
N_{\mathrm{field}}=E_{\mathrm{field}}\int_{0}^{\infty}\,
{\frac{\mathrm{d}\Gamma}{\mathrm{d}T_{\mathrm{E}}}}\, {\cal
E}(T_{\mathrm{E}})\, \mathrm{d}\,T_{\mathrm{E}} \; ,
\end{equation}
where the ``field exposure'' $E_{\mathrm{field}}$ 
is defined in \cite{MACHO00}, as the product
of the number of distinct light curve per field
and the relevant time span. Furthermore, we pay attention
to eliminate the field overlaps.

To characterize the expected timescale we report
the \emph{median} value of the asymmetric distribution 
$(\mathrm{d}\Gamma/\mathrm{d}T_\mathrm{E})_\varepsilon$,
together with the values $T_{\mathrm{E},\,16\,\%}$ and $T_{\mathrm{E},\,84\,\%}$
that single out the extremes of the 68\% probability range around the median.

In Table \ref{tab:nevtot} we report, for both MW and LMC MACHO lens populations,
the total number of the expected events in all the MACHO fields,
in the case of a full MACHO halo, together with the expected timescales.
We recall that the expected timescale varies with the square root of 
the mass of the MACHOs,
and that very short timescales are strongly suppressed by the detection 
efficiency function.

\begin{table}[htb!]
\begin{center}
\begin{tabular}{c|cc|cc}
lens mass&\multicolumn{2}{c}{MW}&\multicolumn{2}{c}{LMC}\\
M$_\odot$ & $T_\mathrm{E}$ (days)&$N_\mathrm{exp}$& 
$T_\mathrm{E}$ (days)& $N_\mathrm{exp}$\\
\hline
$10^{-5}$ &$3.3^{+3.0}_{-1.3}$ & 0.9 & $3.3^{+3.0}_{-1.3}$ & 0.4 \\
$10^{-4}$ &$3.5^{+3.0}_{-1.5}$ & 8.3 & $3.5^{+3.1}_{-1.3}$ & 3.2 \\
$10^{-3}$ &$4.3^{+3.6}_{-1.6}$ & 52  & $5.4^{+4.4}_{-1.9}$ & 13.5 \\
$10^{-2}$ &$8.0^{+6.5}_{-3.1}$ & 115 & $13^{+8.7}_{-4.4}$  & 17.3 \\
0.1       &$20^{+15}_{-8.0}$   & 97  & $36^{+21}_{-13}$    & 9.9 \\
0.2       &$26^{+20}_{-10}$    & 82  & $48^{+27}_{-17}$    & 7.6 \\
0.5       &$41^{+29}_{-16}$    & 59  & $75^{+46}_{-27}$    & 5.0 \\
1         &$55^{+38}_{-21}$    & 44  & $103^{+57}_{-37}$   & 3.4 \\ 
10        &$149^{+79}_{-57}$   & 12  & $245^{+94}_{-87}$   & 0.5 \\
\end{tabular}
\caption{Expected duration, median values with 68\% CL errors,
and expected number of events for a full dark matter halo, respectively averaged
and summed over the MACHO fields, as a function of the MACHO mass.
}
\label{tab:nevtot}
\end{center}
\end{table}

The predicted durations turn out to be almost independent from the position 
for both the halo populations we consider, whereas the issue of the variation of 
the expected timescales with the position across the fields for the 
self-lensing population has been discussed thoroughly in Paper II. 
Indeed, for lenses in the MW halo we find 
a dispersion of the median timescales  towards the different fields smaller 
than $1\%$. In the case of the LMC lenses the dispersion 
is only slightly larger, at most $\sim 5\%$.

The  expected total number of events due to  the LMC stellar halo 
turns out to be $\sim 0.6$, about half of those due to 
the LMC disc-bar self-lensing contribution (Paper II).
The expected median timescale, averaged on the 30 fields,
is $T_{E}=45^{+43}_{-23}$ days.

Overall, we recover the result \citep{MACHO00} that stellar 
lensing alone cannot explain the signal,
so that most of the detected events  must belong either 
to the MW or to the LMC dark matter halo.

\section{The LMC MACHO contribution to microlensing events} 
\label{sec:lmc-res}

A straightforward conclusion to be drawn out of the results
on the expected number of events due to the dark matter MW and LMC halos 
is that, with the implicit hypothesis that the halo fractions 
in both the MW and the LMC halos are the same,
most of the lenses should indeed belong to the MW halo.
Our aim is to challenge this point of view. 

First, we recall the current status about the microlensing
events observed by the MACHO collaboration.
Next, we carry out a statistical analysis
of the observed characteristics of the events
(timescale and spatial distribution).
The purpose is to determine to which extent
the available data allow to distinguish
between the two halo populations.
Eventually, using a likelihood analysis 
based on the microlensing rate,
we study whether a viable solution
to the MACHO puzzle can come
from a significant contribution
of a lens population belonging to the LMC halo.

\subsection{The microlensing MACHO candidates} \label{sec:macho-res}

In the final analysis of 5.7 years of data in 30 fields towards the LMC,
the MACHO group presented two sets of microlensing candidates,
set A and set B, with 13 and 17 candidates respectively, the former being
a subsample of the latter \citep{MACHO00}.

Further works allowed to get more information on some of these candidates.
The lens for the event LMC-5 is located in the Galactic disc \citep{macho-nat01}.
LMC-22 has been identified to be very likely a supernova \citep{alcock2001}.
LMC-23 has been acknowledged to be a probable variable stars \citep{bennett05b}.
LMC-9 is a double lens system with caustic crossing \citep{alcock2000}.
The microlensing candidates LMC-9, LMC-20, LMC-22 and LMC-27 belong 
to the set B only.
For most of the remaining events a further photometric follow-up
allowed to confirm the microlensing origin of the flux variation 
\citep{bennett05b}.

In the present analysis we restrict to an homogeneous set
of Paczy\'nski-like events, therefore we exclude LMC-9. Furthermore,
we do not include the Galactic disc lens population, so that we exclude LMC-5,
as well as  all those candidates whose microlensing origin has been 
put in doubt. Accordingly, in the following we shall consider 
a subset of 13 events taken from the original
larger set B, from which we exclude the candidates LMC-5, LMC-9, LMC-22 and LMC-23.
Furthermore, we have verified that our main conclusions
would not change had we started from the smaller set of 11 events
subsample of the original set A, just excluding  the candidates LMC-5 and LMC-23.

\subsection{Duration and position: a statistical analysis} \label{sec:stat}

As previously noted, the expected timescale distributions
for microlensing events due to lenses either in the LMC or in  the MW halo 
are almost independent
from the position. This property allows us to carry on
an analysis in which we  compare the observed
timescales to the predicted one for each population. In particular
we investigate whether it is possible to draw from such an analysis any conclusion
on the relative fraction of the MW over the LMC dark matter halo events.
Here we neglect the stellar lensing contributions.

\begin{figure}[hbt!]
\resizebox{\hsize}{!}{\includegraphics{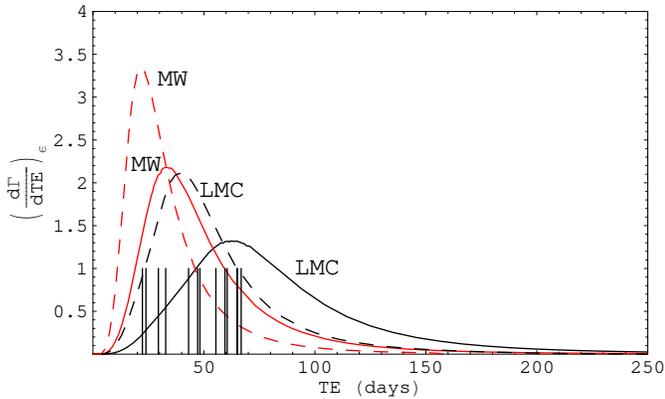}}
\caption{Normalized differential rate $(d\Gamma/dT_{\textrm E})_\varepsilon$
for both MW and LMC halos for 0.2 and 0.5 M$_\odot$, dashed
and solid lines respectively . Superimposed the value of the observed durations.
The $y$-axis values are in $10^{-2}$ units. 
\label{fig:rate-te}}
\end{figure}

\begin{table}[htb!]
\begin{center}
\begin{tabular}{ccccc}
lens mass (M$_\odot$)&$ks_\textrm{MW}$ &$ks_\textrm{LMC}$ &
$ks_\alpha$ & $\alpha$\\
\hline
0.01 & 1.000 &1.000 &1.000 &0.000\\
0.10 & 0.999 &0.683 &0.683 &0.000\\
0.15 & 0.994 &0.159 &0.085 &0.095\\
0.20 & 0.984 &0.450 &0.118 &0.347\\ 
0.22 & 0.969 &0.655 &0.144 &0.471\\
0.30 & 0.877 &0.957 &0.240 &0.703\\
0.40 & 0.579 &0.997 &0.228 &0.871\\
0.50 & 0.176 &1.000 &0.176 &1.000\\
0.60 & 0.398 &1.000 &0.398 &1.000\\
0.80 & 0.743 &1.000 &0.743 &1.000\\
1.00 & 0.911 &1.000 &0.911 &1.000\\
\end{tabular}
\caption{Kolmogorov-Smirnov test results}
\label{tab:ks-alp}
\end{center}
\end{table}

For a given value of the MACHO mass,  the expected LMC
median timescales are larger than the MW ones \citep{gould93}. 
In Figure \ref{fig:rate-te} we show in the same plot the  normalized, differential
rate distribution for lenses in the LMC and the MW halo, in correspondence of
two values of the MACHO mass ($0.2$ and $0.5\, \mathrm{M}_\odot$). 
Superimposed, the vertical lines indicate the Einstein time 
of the observed events.
To further investigate this issue, we make use of the Kolmogorov-Smirnov test 
(hereafter KS). This allows us to test the null hypothesis
that the events are drawn from a given population.
The resulting KS coefficient gives the significance level of the test.
In the first place, we apply the KS hypothesis test  separately 
to the two populations of lenses in the  MW and the  LMC halo.  Then we introduce
a parameter $\alpha$, defined as the ratio of the MW events over the total 
(MW plus LMC events),
in order to explore the possibility that an intermediate solution,
with the two populations mixed, has to be preferred.
To this purpose we look for the value of $\alpha$ that minimises 
the corresponding KS coefficient.

\begin{figure}[hbt!]
\resizebox{\hsize}{!}{\includegraphics[scale=0.6]{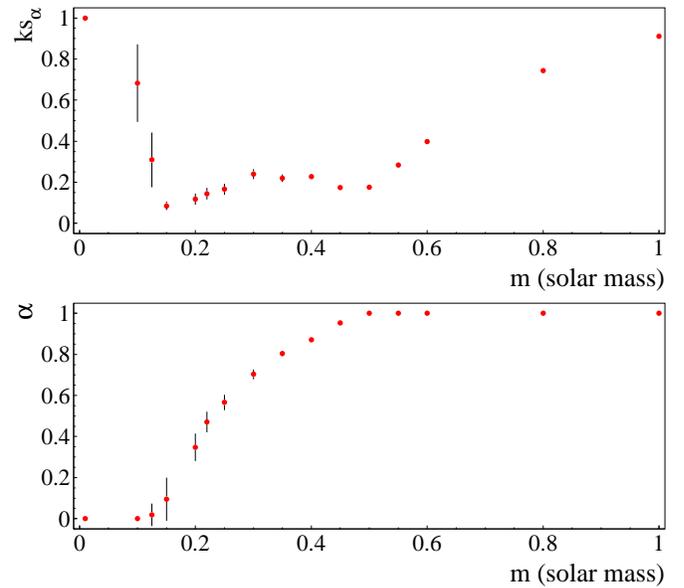}}
\caption{Kolmogorov-Smirnov and $\alpha$ coefficients
as a function of the MACHO mass. $\alpha$ is the ratio
of the MW over the total number of dark matter halo events.
\label{fig:ks-alp}}
\end{figure}

To take into account the variations of the microlensing rate across
the fields, mainly in the case of  lenses in the LMC halo, we carry out 
the test on each of the MACHO fields separately. We then evaluate and report 
the  value averaged on the $30$ MACHO fields. The associated dispersion gives 
the error bars drawn in Figure \ref{fig:ks-alp}.

In Table \ref{tab:ks-alp} and Figure \ref{fig:ks-alp} we present the results
of this analysis. We report the value of the parameter $\alpha$ and
the KS coefficient in the three cases considered,
$ks_\textrm{MW},\,ks_\textrm{LMC}$ and $ks_\alpha$
respectively, as a function of the MACHO mass. We report
the results only in the range 0.01-1 M$_\odot$ where the
preferred solution are found (see below).

When we consider the MW and the LMC halos, separately, the  solutions 
with the highest level of confidence fall in correspondence of  a MACHO mass 
$\approx 0.5$ M$_\odot$ and $\approx 0.15$  M$_\odot$ respectively. 
This result is confirmed by the microlensing 
rate normalised distributions in Fig. \ref{fig:rate-te}, where  the profile 
corresponding to the case of $0.5\,\mathrm{M}_\odot$ MW lenses  is almost 
coincident with that of $0.2\,\mathrm{M}_\odot$ LMC lenses.

For the  case of the combined populations,  we find that the coefficient 
$ks_\alpha$  presents  two  minima,  near to the two values of  
mass found in the previous case. 
The absolute minimum, that we note to be lower than the values obtained 
in the case of the separate test analysis, is found at 
$m=0.15\,\textrm{M}_\odot$ with $\alpha\sim 0.1$,  and gives us 
the parameters with the highest confidence level. Moreover 
we observe that the $\alpha$ parameter grows monotonically  
as a function of the mass from 0 up to 1 (Fig. \ref{fig:ks-alp}).

We conclude that the statistical analysis made 
on the duration of the events 
gives a first suggestion that a significant fraction of the
observed events could belong to the LMC MACHO halo population.

\begin{figure}[hbt!]
\resizebox{\hsize}{!}{\includegraphics[scale=0.6]{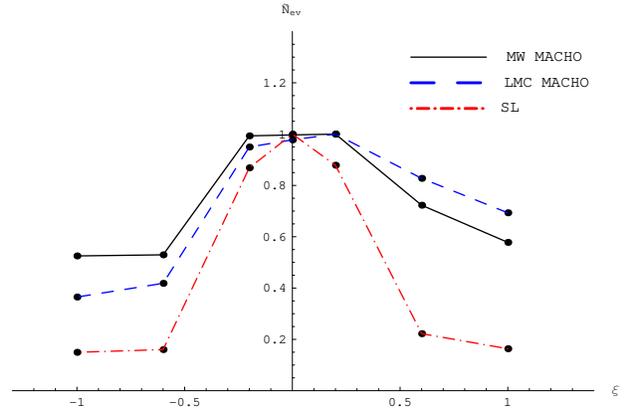}}
\caption{Normalized expected number profiles along the axis,
pointing south-west, orthogonal to the line of nodes
and passing through the LMC centre, for different lens populations:
self lensing, dot dashed line; MW MACHO, solid line;
LMC MACHO, dashed line. Values on the $\xi$ axis are in kpc. 
\label{fig:nnpx}}
\end{figure}

Next we consider the issue of the spatial distribution of the 
observed events \citep{gould93}.
The optical depth profiles clearly show that the LMC halo events are 
characterized, with respect
to LMC self-lensing and to MW halo ones, by a strong
asymmetry  with respect to the line of nodes (Paper II). Looking  
at the expected number of events per field, this asymmetry is somewhat weakened, 
but still present, because of the different source density as a function 
of the position.  In order to get an insight in the more complex 
two dimensional situation we have to deal with,
in Figure \ref{fig:nnpx} we show the normalized number of the expected 
events (${\tilde N}_{\textrm{ev}}$), for the different lens populations 
we consider, evaluated along the axis orthogonal to the line
of nodes passing through the LMC centre (the $\xi$ axis in the plot). 
${\tilde N}_{\textrm{ev}}$ has been calculated in the LMC centre and 
in six further positions,  specularly symmetric two 
by two with respect to the centre. 
This plot shows clearly that  the distribution in the case of self-lensing 
events is  symmetric, and moreover that outside the bar region  
it declines sharply. The profile for MW events presents 
a slight asymmetry with respect to the line 
of nodes, whereas that corresponding to LMC MACHO lenses have a pronounced 
asymmetric distribution. 

\begin{figure}[hbt!]
\resizebox{\hsize}{!}{\includegraphics{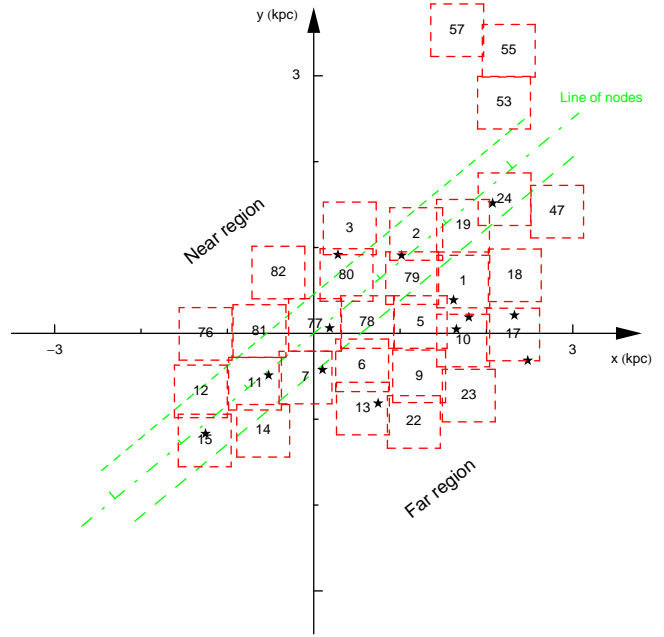}}
\caption{Location of the 30 MACHO fields in a reference frame
centered in the LMC centre with the $x$-axis antiparallel
to the right ascension, the $y$-axis parallel to the declination,
and the $z$-axis (not shown) pointing towards the observer.
The location of the 13 microlensing candidates, subset of the
original set B of MACHO candidates we use
in the present analysis (Section \ref{sec:macho-res}), 
is shown. Also shown, the position
of the line of nodes and the central band around it, which
we exclude in the asymmetry analysis.
\label{fig:map}}
\end{figure}

In Paper II we have addressed the question whether the
observed asymmetrical distribution of the detected events,
that goes indeed in the same sense  predicted
by a halo LMC population, does reflect the observational
strategy, mainly to argue against the self lensing origin
of the events. Here we take advantage of the knowledge
of the expected number of events for all the
populations of interest to further study this issue.

\begin{figure*}[tbh]
\begin{center}
{\includegraphics[scale=0.33]{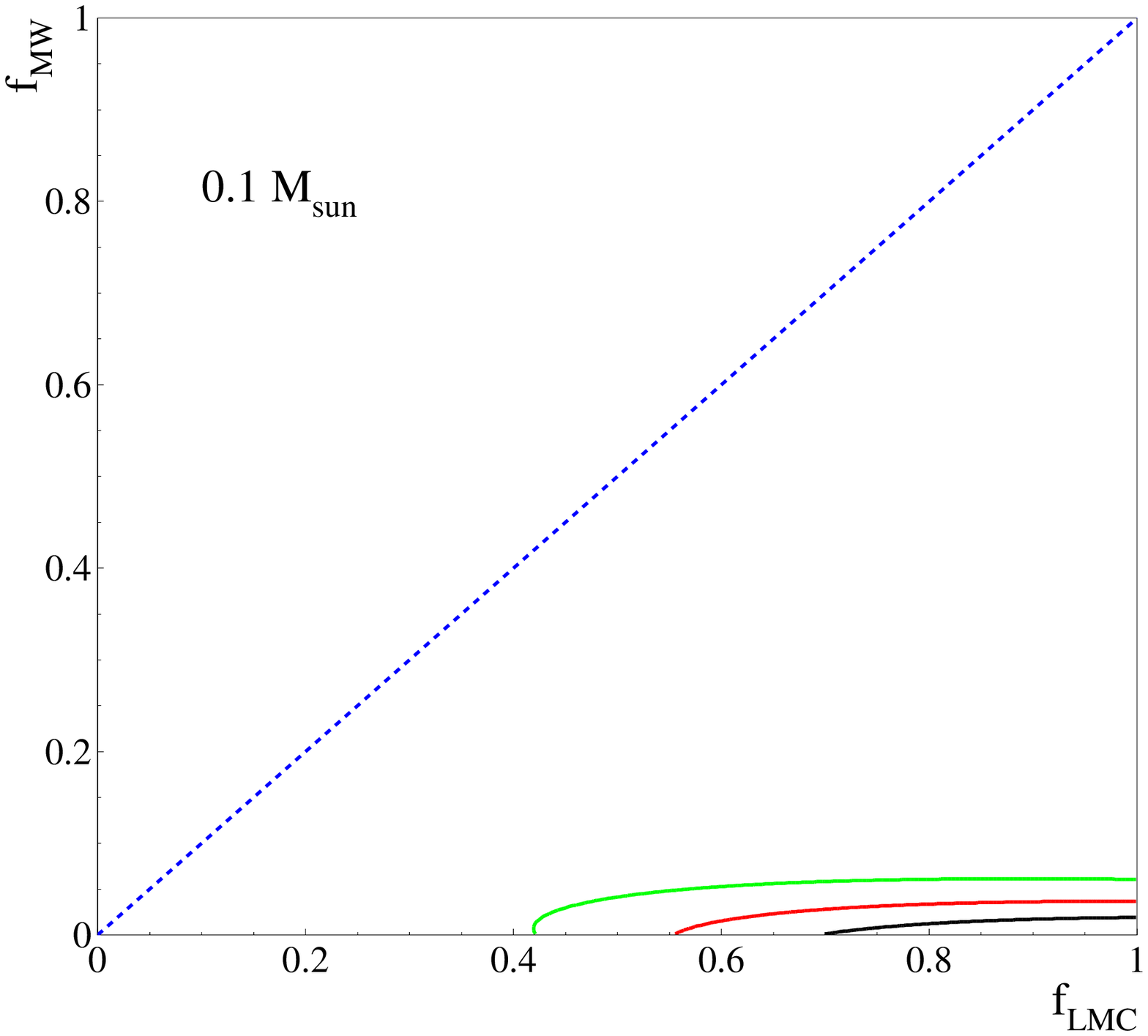}}
{\includegraphics[scale=0.33]{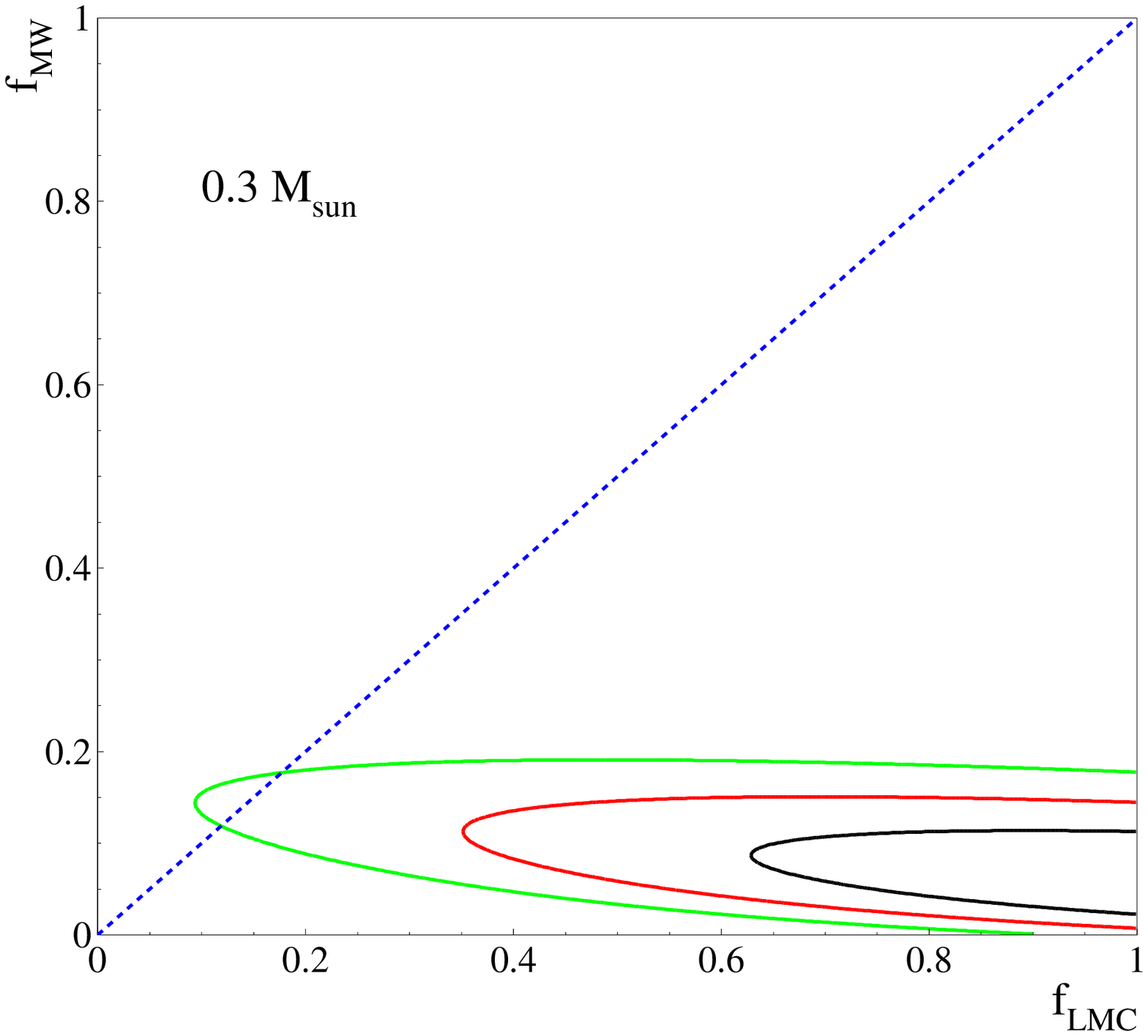}}
{\includegraphics[scale=0.33]{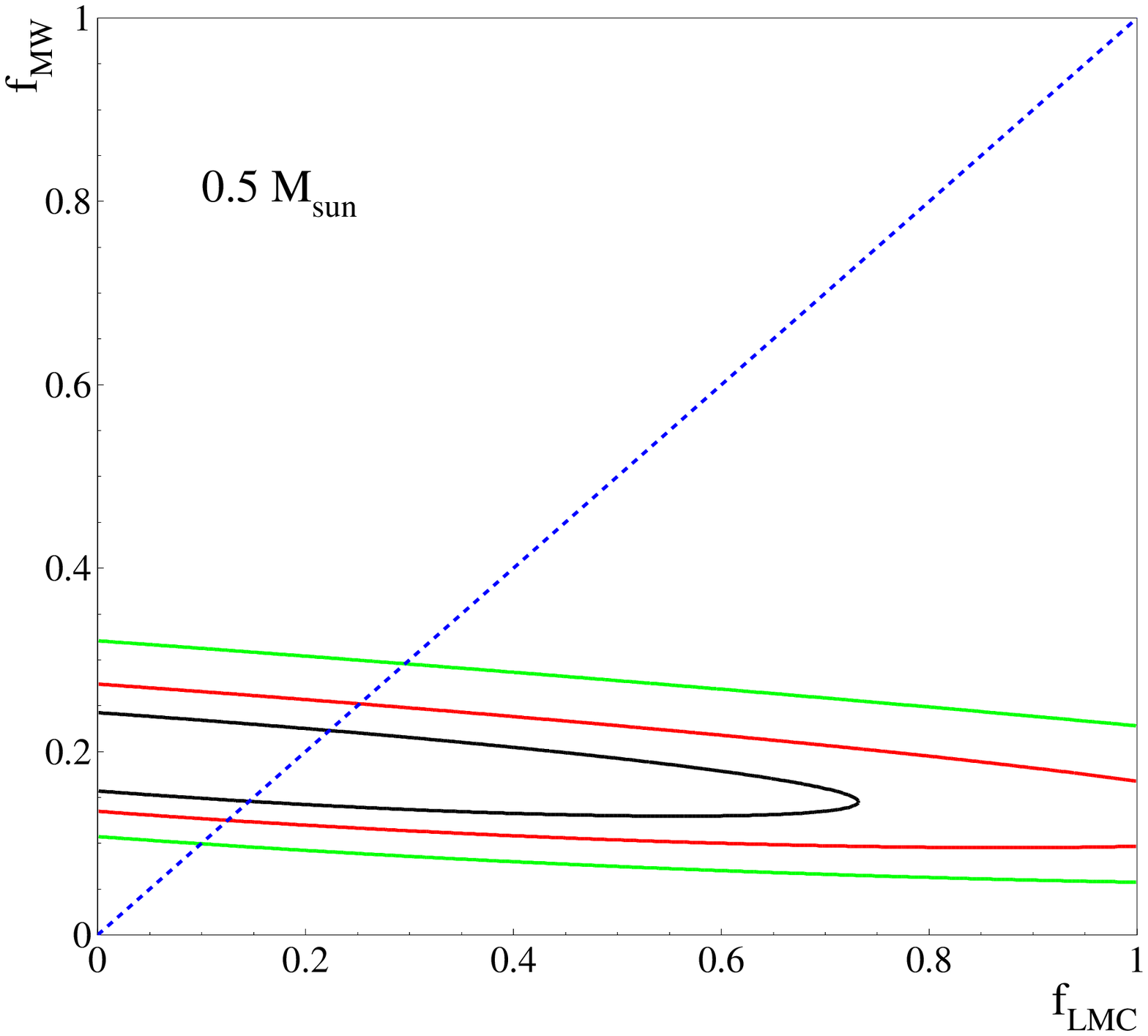}}
{\includegraphics[scale=0.33]{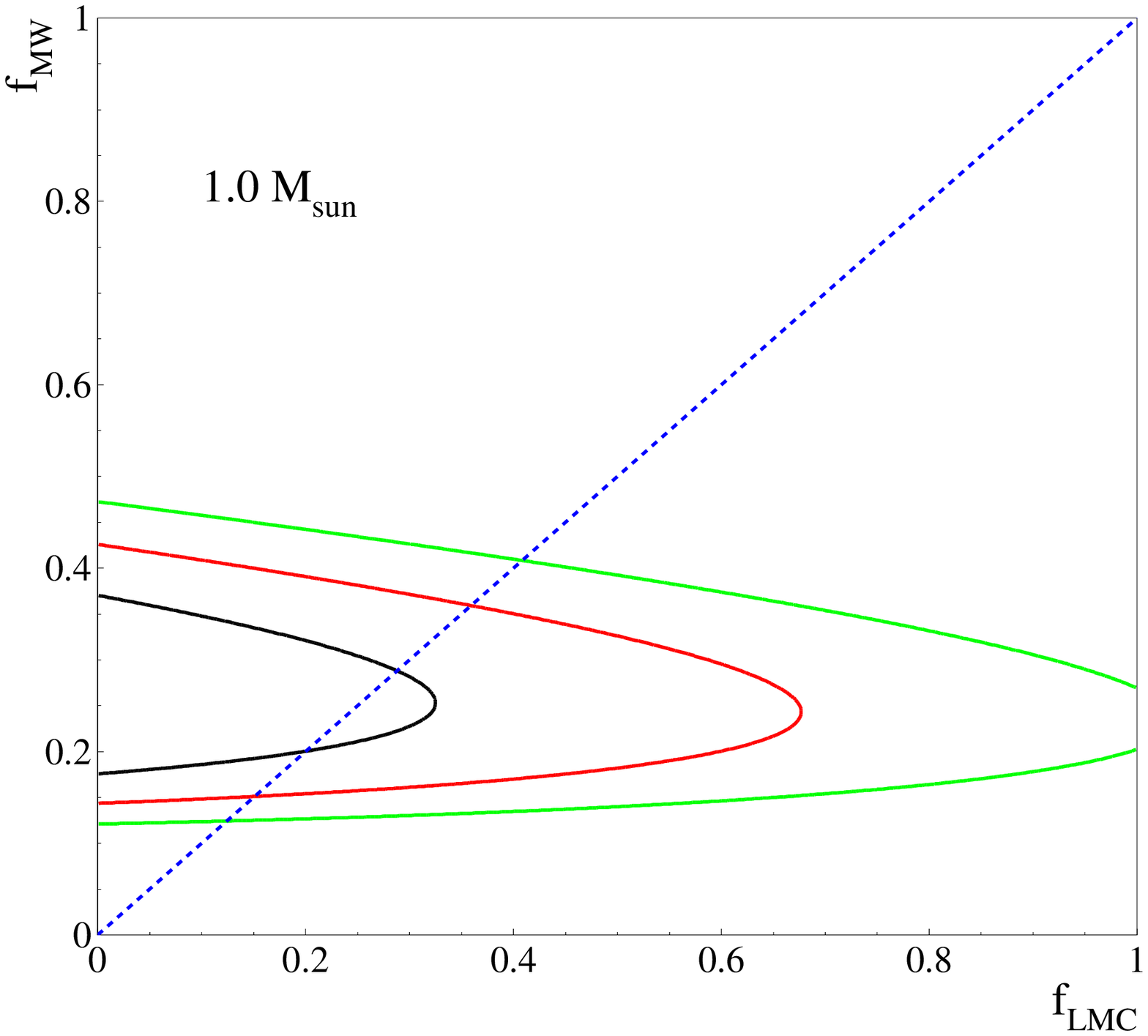}}
\caption{Probability isocontours with 34\%, 68\% and 90\% regions
for the LMC and MW dark matter halo fraction for four values of the MACHO mass.
\label{fig:like3}}
\end{center}
\end{figure*}

We take the line of nodes as the symmetry axis, and  
as in Paper II we bin the observed space in two regions,
the north-east ``near'' and the south-west ``far'' one 
(Fig. \ref{fig:map}).
We  delimit them by two straight lines parallel
to the line of nodes, each at a distance from the latter 
of $\approx 0.35\,\textrm{kpc}$,
the 1 $\sigma$ error in position as calculated by \cite{vdmarel02}.
Note that this way the innermost LMC regions are excluded 
from the asymmetry analysis.
Our purpose is to test the asymmetry with respect to the prediction 
of the different lens populations,
assigning the probability $p_i$ that a microlensing 
event would  fall in the first or second region to be proportional 
to the expected number of events of the given population.  
As in Paper II we make use of the non-parametric Pearson test,
that allows us to evaluate the probability to accept  the null hypothesis 
that the $p_i$ match the measured distribution, for which we get  
1 and 7 observed events in the near and far region,  respectively. 
The analysis is carried out normalizing
the number of expected events of each population to that of the observed events.

The result of this analysis gives us a probability of 46\% and 26\% for a
MACHO LMC and MW halo lens population, respectively.
This outcome puts clearly in evidence the lack of a predominant lens population.
Indeed, the expected smoother spatial distribution of MW halo lenses is only poorly
coherent with the observed distribution, thus challenging the explanation that attributes most
of the events to this population.  Rather, we find a much better agreement
with the expected asymmetric distribution of LMC halo events.
Finally, we remark that, contrary to the previous analysis
based on the timescale distribution,
this result turns out to be independent from the value of the MACHO mass.
As a byproduct of the present analysis (confirming
that carried out in Paper II), we note that
the probability to accept the hypothesis
of a self-lensing origin on the basis of the observed 
spatial distribution turns out to be  only of 19\%.

Both these analysis, carried out working on
\emph{normalized} distributions, i.e. independent from the actual
halo fraction, indicate that  a large fraction of the lenses
could indeed belong to the LMC dark matter halo. 

\subsection{MW and LMC: two different halo fractions?} \label{sec:results}

The previous analysis provided us with two important clues both showing that a significant 
fraction of the events detected by the MACHO collaboration
could be part of the LMC dark matter halo.  
In order to reconcile this result with the predicted number 
of events (Table \ref{tab:nevtot})
we now  drop the hypothesis of equal  halo MACHO  fractions, 
in the MW and the LMC halo.

We start by evaluating the likelihood function 
\begin{eqnarray}\label{eq:likelihood-mf}
&&L\left(f_{\textrm{MW}},\,f_{\textrm{LMC}}\right) =\nonumber\\ 
&&\exp\left(-N_\mathrm{exp}\right)\,
\prod_{i=1}^{N_\mathrm{obs}}\left[ 
E\, {\cal E} (T_{\mathrm{E}_i})\, \frac{\mathrm{d}\Gamma}
{\mathrm{d}T_{\mathrm{E}}}\,\left(T_{\mathrm{E}_i}\right)
\right]\,,
\end{eqnarray}
where $f_{\textrm{MW}},\,f_{\textrm{LMC}}$ are the halo fractions
for the MW and the LMC respectively.
For both $N_\mathrm{exp}$, the expected number of events,
and the differential rate $\frac{\mathrm{d}\Gamma}{\mathrm{d}T_{\mathrm{E}}}$,
we sum over all the lens populations (including the stellar ones), multiplying
the MACHO contributions for the appropriate halo fraction.
The product runs over the $N_\mathrm{obs}$ observed events.
By Bayesian inversion, we use a flat prior probability,  
it is then possible to compute the probability
distribution for the halo fractions given the observed events,
$P\left(f_{\textrm{MW}},\,f_{\textrm{LMC}}\right)$.
Note that we are now taking the MACHO mass as a parameter, equal for 
both halo populations.

\begin{figure}[hbt!]
\resizebox{\hsize}{!}{\includegraphics[scale=0.6]{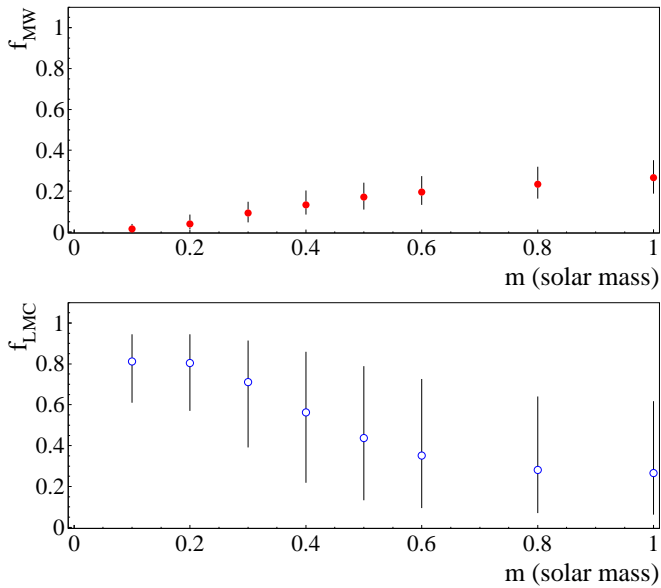}}
\caption{MW and LMC dark matter halo fraction, median value
with 68\% CL errors, as a function of the MACHO mass.
\label{fig:like3m}}
\end{figure}

In Figure \ref{fig:like3} we show, for four 
values of the MACHO mass, the 2-dimensional probability isocontour
for the two halo fractions. Eventually, after marginalisation 
over one variable with respect to the other, we get the results 
for  the two halo fractions as a function of the MACHO mass
shown in Figure \ref{fig:like3m}.

The main outcome of this analysis is that
in a significant range of the MACHO mass, the LMC and MW
dark matter halo fractions are \emph{not} expected to be equal.
In particular, from  0.1 up to 0.3 M$_\odot$,
both a large value for $f_{\textrm{LMC}}$
and a small one for $f_{\textrm{MW}}$ are expected.

This behaviour is mainly due to the observed timescales.
It is therefore not surprising that this result
is coherent with that obtained with the KS test,
where we have found a preferred value of the mass of about $0.2\,\textrm{M}_\odot$,
with a significant expected contribution from LMC MACHO halo lenses.
The likelihood analysis gives, for $m=0.2\,\textrm{M}_\odot$,
$f_\textrm{MW}\sim 4\%$ and $f_\textrm{LMC}\sim 80\%$. 
At face value, given the number of expected events,
we get to the overall prediction
of about 6-7 events to be attributed to the 
LMC dark matter halo lens population,
2-3 to the MW halo one, to be looked for
among those with the shorter durations, 
still allowing for a contribution, about 2-3 events,
distributed among self-lensing and LMC stellar halo lenses.

For larger values of the mass, the LMC MACHO halo fraction
turns out to be almost degenerate though compatible with zero, and,
for 0.5 M$_\odot$, we recover the result of a MW MACHO
halo fraction of about 20\% \citep{MACHO00}. This is again coherent
with the issue of the KS test.

We stress that the outcome of this analysis has to be looked at together
with the outcomes of the previous analysis on the timescale and the spatial distributions 
of the observed events. Overall, they indicate that $\left.i\right)$ a sizeable fraction 
of the observed events shows characteristics in accord with those
expected for a MACHO LMC halo population; $\left.ii\right)$ 
such a contribution may be expected by dropping the hypothesis 
that the halo fractions in form of MACHOs in the MW and the LMC dark matter halo are equal.

As a last point we discuss the dependence of our results from the choice
of the LMC dark matter halo parameters, the central density and the truncation radius  (Sect. \ref{sec:models}).
We find that any variation of the parameters in a reasonable range
around their fiducial values do not change, at least qualitatively, our results.
As for the central density, any variation
downward (upward) is linearly related with a
corresponding change in the total number of expected
events. This implies a scaling respectively upward (downward)
for the halo fraction without affecting the main conclusion
on the contribution of the LMC halo.
As for the truncation radius ($R_t$), in first approximation
the situation is similar, as, roughly, a smaller (larger) value of $R_t$ give rises to
a smaller (larger) total LMC mass. However, the issue is slightly more subtle.
First, the problem is not symmetric with respect to the choice
of our fiducial value, $R_t=15$ kpc. Second, 
different choices for $R_t$ enhance different spatial distribution
for the number of expected events.
In particular it results that:
$\left.i\right)$ the decrease in the number of expected events, for values of $R_t$
smaller than the fiducial value, is relatively larger than the increase obtained
by choosing larger values;  $\left.ii\right)$ the spatial asymmetry 
of MACHO LMC halo events is enhanced for smaller value of $R_t$. 
(Both effects are easily explained as most of the lenses
are expected to be located in the innermost LMC region.)
We have tested our results
with 4 values of $R_t$, respectively smaller (larger) of our
fiducial value by 20\% and 40\%. 
The relative average decrease in the number of expected
events for $R_t=9$ and 12 kpc with respect to the $R_t=15$ kpc case
is $\sim 20\%$ and 8.7\%, whereas the relative increase for $R_t=18$ and 21 kpc
is $\sim 7.4\%$ and 14\% respectively. As for the spatial
distribution, we characterise the asymmetry by the relative difference
between the number of expected events
evaluated at the extremes of the $\xi$ axis
(as defined in Fig. \ref{fig:nnpx}) $\xi=1$ kpc and $\xi=-1$ kpc.
With respect to the fiducial case, where this turns out to be of $47\%$,
for $R_t=9$ and 12 kpc we find a relative increase of 11\% and 4\%, whereas
for $R_t=18$ and 21 kpc a relative decrease of 2\% and 6\%, respectively.
Overall, smaller values of $R_t$ strengthen our conclusions.

\section{Conclusions} \label{sec:the-end}

In this paper we have addressed the issue
of the interpretation of the microlensing results
toward the LMC. In particular, starting from
the sample of microlensing candidates reported
by the MACHO collaboration, we have discussed the contingent
contribution of a lens component belonging to 
the LMC dark matter halo besides that of the MW one.
As a main result of the present analysis we show that
a sizeable fraction of the observed events,  up to about half of the total,
could indeed be part of the first component.

We summarise our analysis as follows. First, we have compared the observed timescales
with those expected for the two different MACHO populations, the MW and the LMC one.
As a result we have shown that the preferred values for the MACHO mass are about 0.5 and 0.2 M$_\odot$
respectively and, through a KS test,  that the latter solution is preferred.
Second, we have studied the spatial distribution of the observed events, 
recalling that, because of the inclination of the LMC disc
with respect to our line of sight, an asymmetry is expected for LMC halo events. As a result we have shown
that, independently from the value of the MACHO mass,  the observed distribution matches
better that expected for a LMC halo population with respect to that of a MW halo population.
Overall, these are clues suggesting the presence of a significant 
MACHO LMC halo population among the observed events.

The extremely larger value of the overall MW halo mass with respect to the LMC one 
implies that generally one can safely ignore the LMC halo component.
In order to explain such a large contribution of the latter, 
one way out is to consider that the halo fractions in form of MACHOs of the two components,
the MW and the LMC halos, may be different. Coherently with the timescale analysis, this issue turns out to be
strongly dependent on the value of the MACHO mass.

In order to get to more quantitative results we have tested this hypothesis through  
a likelihood analysis. Eventually we have shown that for a large range of mass values
a different (and larger) fraction for the LMC halo with respect to the MW one is indeed expected.
In particular, for MACHO mass of $\sim 0.2\,\textrm{M}_\odot$
we evaluate a high halo fraction for the LMC, $\sim 80\%$,
together with a small one for the MW, $\la 5\%$,
thus implying that about half of the observed events
should belong to the LMC dark matter halo.
On the other hand, for MACHOs of $\sim 0.5\,\textrm{M}_\odot$
we recover the well known result of a MW halo fraction $\sim 20\%$
with a (possibly) negligible contribution from the LMC dark matter halo.

A possible explanation to the origin of different halo fractions
could come from the different formation history of the two galaxies,
or, more simply, could be related to the fact that one observes
all the LMC halo but, practically, only a line of sight through the Galactic halo.

These conclusions should be taken
\emph{cum grano salis}. The overall implicit
assumption is the validity of the MACHO results, whereas they are actually
challenged by the EROS collaboration. An intrinsic limit is then due to the
small statistic at disposal. The SuperMACHO collaboration \citep{becker04}
is expected to provide a larger sample of candidates
spread over a much larger field of view and this should allow
to put firmer constraints on this problem. Furthermore, the model issue,
in particular for the LMC components, is still a matter of debate. 
Our analysis shows, however, that it is in principle possible to characterize
and distinguish the two halo lens populations and,
moreover, challenge
the usual implicit assumption of equal halo fractions
in form of MACHO for both the MW the LMC dark matter halos.

\begin{acknowledgements}
We thank Andy Gould for useful comments and discussions.
SCN was partly supported by the Swiss National Science Foundation.
FDL work was performed under the auspices of the EU, that has provided
financial support to the "Dottorato di Ricerca Internazionale in
Fisica della Gravitazione ed Astrofisica" of the Salerno University,
through "Fondo Sociale Europeo, Misura III.4".
GS acknowledges support for this work provided by MIUR through PRIN 2004
"Astroparticle Physics", protocol number 2004024710\_006, 
and by research funds of the Salerno University.
\end{acknowledgements}

\bibliographystyle{aa}
\bibliography{new}
\end{document}